\documentclass{ws-ijmpa}
\usepackage[super,compress]{cite}
\usepackage{graphicx,xcolor}

\usepackage{amsmath,amssymb}
\usepackage{multirow,enumitem}
\usepackage{diagbox}
\usepackage{array}
\usepackage{hyperref}

\begin{document}
\markboth{Yu-Cheng Qiu, Hui-Yu Zhu}{Probing Neutrino Sector via A Statistical Approach}

%
\catchline{}{}{}{}{}
%

\title{Probing Neutrino Sector via A Statistical Approach}

\author{Yu-Cheng Qiu}

\address{Department of Physics, Hong Kong University of Science and Technology\\
Hong Kong S.A.R., China\\
yqiuai@connect.ust.hk}

\author{Hui-Yu Zhu}

\address{Department of Physics, Hong Kong University of Science and Technology\\
Hong Kong S.A.R., China\\
hzhuav@connect.ust.hk}

\maketitle


\begin{abstract}
We apply the idea of landscape (motivated by string theory) to study the statistical nature of parameters/couplings in the standard model of strong and electroweak interactions. Following the success of this approach on the fermion masses,  we discuss the divergent behavior of the probability distributions of other physical parameters/couplings to obtain some insights on the quantities that cannot be measured by current experiments but can be relevant in cosmology, in particular those in the neutrino sector. From a relatively strongly mixed PMNS matrix, we argue that the probability distribution of heavy neutrino mass $P(M)$ does not diverge at $M=0$. This analysis favors the degenerate heavy neutrino scenarios.

\end{abstract}


\ccode{PACS numbers: 12.15.Ff, 14.60.Pq.}


	\section{Introduction}
	
	The neutrino oscillation measured by experiments in 1998~\cite{Super-Kamiokande:1998kpq} shows clearly that the Standard Model (SM) of particle physics is inadequate, which poses two major questions on the neutrino sector: why the masses are so small and why they oscillate. The Seesaw mechanism is proposed to explain the smallness of neutrino masses by introducing a heavy sector to suppress them. The origin of oscillation is reviewed recently~\cite{Kupiainen:2021agu}. More detailed reviews and outlook on the neutrino could be found in Ref.~\citen{Athar:2021xsd,Senjanovic:2020rcq,Choubey:2021bln}. Similar to the CKM matrix for quarks, neutrinos have their mixing matrix denoted as the PMNS matrix. Due to the possible Majorana nature of the neutrinos, there are six real parameters in the PMNS matrix, consisting of three mixing angles $\theta_{\rm p}$s, one Dirac phase $\delta_{\rm p}$ and two Majorana phases $(\alpha_{\rm M}, \beta_{\rm M})$. Neutrino oscillation could be determined by $\theta_{\rm p}$s and $\delta_{\rm p}$, where Majorana phases play no role. Neutrinoless double beta decay ($0\nu\beta\beta$) is a possible way to detect them if the light neutrinos are Majorana fermions and the strength is affected by $\alpha_{\rm M}$ and $\beta_{\rm M}$.
	
	Matter-antimatter asymmetry of the universe is a long-standing puzzle of cosmology. Leptogenesis is a promising solution~\cite{Fukugita:1986hr,Davidson:2008bu}. The simplest scenario contains heavy Majorana neutrinos whose CP-violating decays violate the lepton number. Then, with the help of the electroweak baryon number violating sphaleron mechanism, the lepton number violation transfers to baryon number violation and explains the baryon-asymmetry of our universe. The involvement of the heavy neutrino sector introduces extra parameters into the effective theory, so a quantitative detailed description is a bit involved. 
	
\begin{figure}
\centering
\includegraphics[scale=0.9]{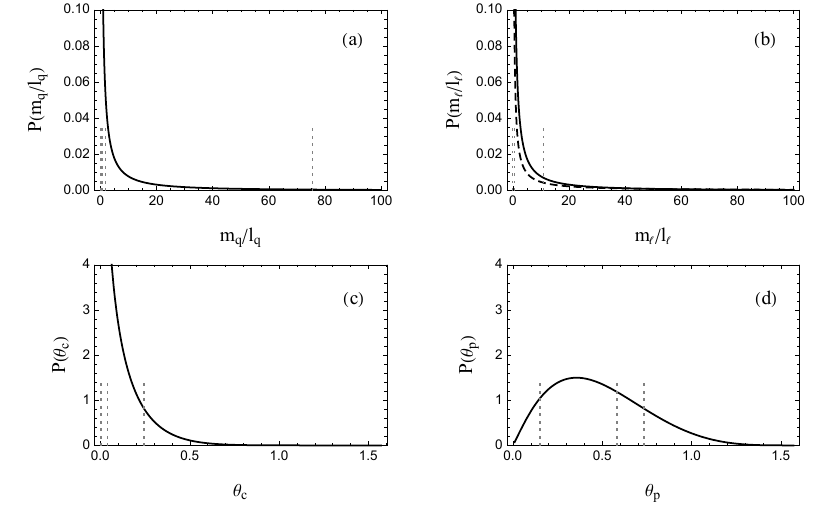}
\caption{(a) and (b) are fitted mass distributions for quarks and charged leptons, where dotted-vertical gray lines label rescaled quark and charged lepton mass values. Here we use current quark masses~\cite{ParticleDataGroup:2020ssz}, which are $\{2.3,4.8,95,1275,4180,173210\}$ MeV and masses for the charged leptons are $\{0.511,106,1777\}$ MeV. Overall scales are $l_{\rm q}=2290\,{\rm MeV}$ and $l_\ell=164\,{\rm MeV}$. (c) and (d) are fitted distributions for CKM and PMNS mixing angles in Standard Parametrization. Solid-vertical lines are mixing angles in rad according to table~\ref{tab:data}. Here we take $\mathcal{NH}$ for $\theta_{\rm p}$ as an illustration since the difference of mixing angles between $\mathcal{NH}$ and $\mathcal{IH}$ is small. The dashed curve in (b) labels the light neutrino mass distribution with type I seesaw mechanism, which is more diverged than that of quarks and charged leptons. The overall mass scale of light neutrinos is also much smaller than that of charged leptons, $l_\nu\ll l_\ell$.}
\label{fig:dist}
\end{figure}

	SM with neutrino masses included is a low energy four-dimensional effective theory consisting of $25$ free parameters, (including $13$ masses, three gauge couplings, one weak mixing angles, four CKM parameters and four neutrino oscillation parameters), which could only be determined by experiments. Reference~\cite{Andriolo:2019gcb} finds that the statistical  distribution for both quark and charged lepton masses fits the same divergent behavior as mass $m \to 0$ :
\[P(m) \to m^{-1+k}, \quad \quad k=0.269\;,\]
even though the mass scales for the quarks and the charged leptons are very different. With the seesaw mechanism for neutrinos, one calculates that $k_{\nu} =k/2=0.134$ for the light neutrino masses, which fits the data quite well; so one can push forward and use this method to study other parameters in SM. This work is an attempt to adopt the statistical approach to analyze the relation between the physical parameters. Here we apply the bottom-up approach that uses a known model distribution (like Weibull distribution for fermion masses in Ref.~\citen{Andriolo:2019gcb}) to fit the measured data, which gives some key information about their intrinsic distributions. Fitted distributions are shown in figure~\ref{fig:dist}. Together with relations between the parameters, we analyze the distribution of the heavy neutrino sector.

	The PMNS and CKM matrices have some similarities. CKM matrix is a unitary three-by-three matrix with four real degrees of freedom (DOF) in it, which are usually parametrized as three mixing angles $\theta_{\rm c}$s and one CP-violating phase $\delta_{\rm c}$ in standard parametrization (SP). Mixing angles do not violate CP. Similarly, the $\theta_{\rm p}$s in the PMNS matrix do not violate CP invariance while the phases $(\delta_{\rm p},\alpha_{\rm M},\beta_{\rm M})$ do. This indicates that the mixing angles and complex phases are sampled from different probability distributions. Also, under the SP~\cite{ParticleDataGroup:2020ssz}, values of $\theta_{\rm p}$s are statistically relatively larger than $\theta_{\rm c}$s, which are locally distributed around zero as indicated in table~\ref{tab:data}. Here we treat $\theta_{\rm p}$ and $\theta_{\rm c}$ differently and consider that they are sampled from two distinct statistical distributions. Assuming they follow Beta distribution with different shape parameters, the best fit gives $P(\theta_{\rm c}\to 0)\propto\theta_{\rm c}^{-1+0.58}$ and $P(\theta_{\rm p}\to 0)\propto\theta_{\rm p}^{-1+2.1}$. This shows that the probability distribution $P(\theta_{\rm c})$ diverges at zero while $P(\theta_{\rm p})$ is smooth at the origin as shown in figure~\ref{fig:dist}.
	
\begin{figure}
\centering
\includegraphics[width=8.6cm]{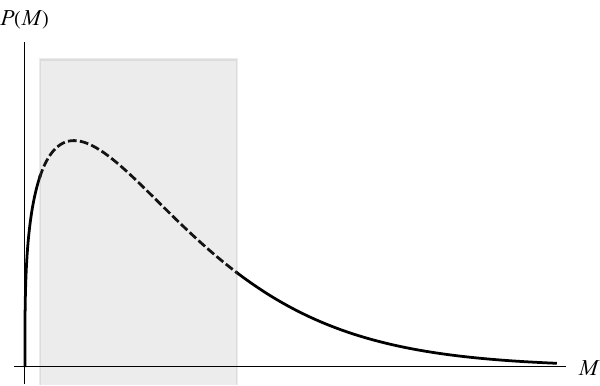}
\caption{Estimate on $P(M)$. The behavior around the origin is determined by divergent power and tail suppression is required by normalization. Gray region is unknown and dashed curve is the reasonable guess without non-trivial form.}
\label{fig:main}
\end{figure}

$P(\theta_{\rm p})$ contains contributions from heavy neutrinos in type I seesaw mechanism and the divergent power of it encodes the information of the statistical distribution of the heavy neutrino masses. In a simple set-up, this leads to the $P(M\to 0)\propto M^{-1+k_{\rm M}}$ where $k_{\rm M}\gtrsim 1$. Based on this, one argues that the heavy neutrinos are not localized around the origin, and the distribution is sketched in figure~\ref{fig:main}. We find that the probability of $M_3> 20 M_1$ is less than $10\%$ for $k_{\rm M}=1.3$, where $M_3$ is the heaviest mass and $M_1$ is the lightest in the heavy neutrino sector. This means that in the simplest type I seesaw scenario, leptogenesis should consider the decays from all three heavy neutrinos instead of simply $M_3\gg M_2\gg M_1 $, which is the main result of our work.

	This paper is organized as the following. In section~\ref{subsec:bg}, the background and statistical approach are reviewed, together with discussions on the different types of seesaw models. In section~\ref{subsec:review}, we briefly review the similarities and differences between CKM and PMNS matrix, after which we investigate an $O(2)$ toy model which captures the key feature of generation mixing in section~\ref{sec:o2}. The implication on the heavy neutrino sector is explained in section~\ref{sec:k}. Section~\ref{sec:diffp} shows the different parametrization of the mixing matrix would essentially lead to the same conclusion on heavy neutrino mass pattern. Some discussions are laid out in section~\ref{sec:discussion} and section~\ref{sec:summary} is the summary. \ref{appendix:stat} gives a few examples of algebraic-relation-induced relations between divergent powers. \ref{appendix:mc_sim} is the Monte-Carlo  simulation of matrix decomposition and rotation combination, which serves as supporting evidence of some arguments we made.
	
\begin{table}[ph]
\tbl{Experimentally measured mixing angles and Dirac phase under the Standard Parametrization with $1\sigma$ range implied. We adopt the parameter range with SK atmospheric data included. }
{\begin{tabular}{@{}cccccc@{}}
\toprule
	& &$\theta_{12}/^{\circ}$&$\theta_{23}/^{\circ}$&$\theta_{13}/^{\circ}$&$\delta_{13}/^{\circ}$\\
\colrule
CKM~\cite{Charles:2004jd,Hocker:2001xe,Dubois-Felsmann:2003jim,UTfit:2005ras,UTfit:2007eik} & & $13.09^{+0.28}_{-0.28}$ & $2.32^{+0.048}_{-0.035}$ & $0.21^{+0.006}_{-0.005}$ & $68.53^{+2.58}_{-2.46}$\\
\colrule
PMNS~\cite{Esteban:2020cvm,Capozzi:2013csa,ParticleDataGroup:2020ssz} & {$\mathcal{NH}$} & $33.45^{+0.77}_{-0.75}$ & $42.1^{+1.1}_{-0.9}$ & $8.62^{+0.12}_{-0.11}$ & $230^{+36}_{-25}$ \\
	& {$\mathcal{IH}$} &$33.45^{+0.78}_{-0.75}$ & $49.0^{+0.9}_{-1.3}$ & $8.61^{+0.14}_{-0.12}$ & $278^{+22}_{-30}$\\
\botrule
\end{tabular} \label{tab:data}}
\end{table}
	
	\section{Method and Review}
	\label{sec:m_and_r}
	
	\subsection{Background}
	\label{subsec:bg}
	
The CKM matrix is considered as weak-mixing and this pattern is assumed to be protected by some flavor symmetries. Meanwhile, Ref.~\citen{Hall:1999sn} shows the possibility of random structure for neutrino mass matrix, and therefore so does the PMNS matrix. Here we adopt the philosophy that both CKM and PMNS matrices are randomly generated and they have their own statistical properties. In fact, Ref.~\citen{Donoghue:1997rn,Donoghue:2005cf,Donoghue:2009me} have already considered distributions of quark/lepton masses and CKM matrix elements, in which they proposed a (nearly) scale-invariant distribution for mass parameters. Starting from that and by feeding in the data, they explored and successfully demonstrated the naturalness of Standard Model parameters. Here we use an extensive approach. Instead of investigating the whole distribution for physical parameters, we use a model distribution to fit experimental data and focus on the divergent behavior at the origin of them, to probe the extension of SM, which is heavy neutrinos in this work.
	
	The statistical approach is based on an effective landscape of physical parameter space inspired by string theory. String theory naturally introduces a landscape through flux compactification, where various vacua (solutions) are located at. Each solution is a particular choice of discrete flux values, which corresponds to a certain type of geometry and compactification. We start with a region where it is assumed that SM locates in it and this patch is described by potential $V(a_i,\phi_j)$ in the effective field theory. The potential contains the flux parameters $a_i$ and, for a specific flux choice, the moduli fields $\phi_j$ are stabilized through dynamics. In principle, flux values are discretized. When the steps between adjacent values are small enough, $a_i$ could be treated as continuous parameters. If we focus on this specific patch of landscape (with SM inside), it is convenient to consider $a_i$ as random variables sampled from some intrinsic probability distribution (probability density function) $P(a_i)$. String theory has no scales other than string scale $M_{\rm S}$, which indicates that one could simply consider $P(a_i)$s are flat (or smooth) over a range. Physical parameters such as masses or couplings are determined by $a_i$ in the effective potential. By sweeping through the $a_i$ in the landscape, one would generate the effective landscape, where physical parameters have their own statistical distributions.
	
	This statistical approach has been applied to explain the smallness of the observed cosmological constant (CC). By sweeping through flux parameters , it is found that in the racetrack K\"ahler Uplift (RKU) model, $P(\Lambda)$ diverges at $\Lambda=0$~\cite{Sumitomo:2013vla}, which gives a good interpretation on exponentially small observed CC $\Lambda_{\rm obs}\approx 10^{-120}\;M_{\rm Pl}$. Here $M_{\rm Pl}$ is the reduced Planck mass. The electroweak scale $\mathbf{m}\approx 100\,{\rm GeV}$ could emerge by feeding $\Lambda_{\rm obs}$ in $P(\Lambda)$~\cite{Andriolo:2018dee}. Furthermore, with the help of this peaking distribution of $\Lambda$ in the model combining RKU and anti-D3-brane~\cite{Qiu:2020los}, the constraint for other scalar parts in the potential turns out to be very restrictive. In Ref.~\citen{Andriolo:2019gcb}, authors found that it is a good fit for quark mass following a distribution which peaks at zero as $P(m_{\rm q}\to 0)\propto m_{\rm q}^{-1+0.269}$. Since quark and lepton masses are generated from the same type of interaction, namely Yukawa interaction, it is natural to assume that they produce the same peaking behavior. By forcing $P(m_\ell\to 0)\propto m_\ell^{-1+0.269}$ and allowing the overall scale to vary, one could also find an excellent fit to charged lepton mass as shown in figure~\ref{fig:dist}. The different overall scales are reasonable since that quarks couple to QCD and charged leptons do not. It is assumed here that the divergent behavior is from UV theory, which could in principle be calculated if one has the complete knowledge of flux compactification. Using a detailed distribution model, authors in Ref.~\citen{Donoghue:1997rn,Donoghue:2005cf,Donoghue:2009me} already found that peaking distribution is common in the quark and lepton sector. It is intriguing to assume that this peaking behavior is a signature of the string theory that SM has. 
	
	Not all physical parameters are directly generated from the discrete flux parameters. The value of a physical quantity is determined by the dynamics and its value is a function of the flux parameters; so its probability distribution follows from the probability distributions of the flux parameters. Despite the possible non-trivial form, probability distribution of a physical parameter $\mathcal{Q}$ is either smooth or peaked at $\mathcal{Q}=0$, which could be described as
	\begin{equation}
		P(\mathcal{Q}\to 0)\sim \mathcal{N} \mathcal{Q}^{-1+k}+\cdots\;,
		\label{eq:qk}
	\end{equation}
where $\mathcal{N}$ is the normalization constant. The parameter $k$ is denoted as the divergent power at the origin. If $0<k<1$, $P(\mathcal{Q})$ diverges when $\mathcal{Q}\to 0$ (properly normalized distributions do not allow $k<0$). The divergence is absent if $k\geq 1$. There may exist other divergent behaviors like $(-\ln\mathcal{Q})^{k'}$~\cite{Qiu:2020los}, which is weaker than the power divergence and we shall ignore them here. This divergent power $k$ is independent of the scale, which is useful when investigating the link between two different sectors (e.g., a heavy sector and a light sector, which have distinct scales).  We shall focus on this quantity, divergent power $k$, to investigate the (beyond) SM parameters. 
	
The simplest smooth probability distribution is a flat distribution over a range of $\mathcal{Q}$. For ease of comparison, we shall also use a Weibull distribution for some unbounded $\mathcal{Q}$. If the range of $\mathcal{Q}$ is finite, like an angle or a phase, we shall use a (transformed) Beta distribution.	

	Dynamics are usually implied by algebraic relations between physical parameters. There are two kinds of relations between parameters that should be clarified here.
\begin{itemize}
\item First kind is the algebraic relation between parameters, which could be simply expressed as a mapping between them. For example, CC is a function of parameters in the model set-up $\Lambda(c_i)$ or quark mass could be assumed as function of flux parameters $m_{\rm q}(a_i)$.
	
\item The second kind is defined as that parameters are sampled from the same probability distribution. For example, one could consider three mixing angles $\theta_{\rm c}$s in CKM matrix or quark masses $m_{\rm q}$s, which could be treated as independently sampled from the same probability distribution.
\end{itemize}

	If parameter $y$ is determined by parameters $\vec{x}=\{x_1,x_2,\cdots,x_N\}$ through function $y=f(\vec{x})$, the probability distribution of $y$ is determined from $f$ and $P_i(x_i)$. Suppose all elements in $\vec{x}$ are independently sampled, the $P(y)$ could be calculated by
	\begin{equation}
		P(y)=\int \prod_i^{N}{\rm d}x_i P_i(x_i)\delta\left(y-f(\vec{x})\right)\overset{y\to0}{\sim}\mathcal{N} y ^{-1+k_y}+\cdots\;,
		\label{eq:py}
	\end{equation}
where $\delta(z)$ is the Dirac delta function. If the $x_i$s are not independent, above formula does not apply. The divergent power $k_y$ is generally determined by $P_i$ and $f$, which are the input and the model set-up respectively.
	
	The target of this work is the neutrino. The long-standing question that why the neutrino sector is relatively strongly mixed while the quark sector is weakly mixed could be explained by investigating the divergent power of the mixing angle's distribution. Intuitively, the origin of the neutrino mass matrix is different from that of quark mass, if the neutrino is Majorana fermion and the Seesaw mechanism is introduced. It is reasonable that they are sampled from distributions with their own divergent powers. Different scenarios would lead to their own explanation and constraints.

\begin{enumerate}
	
\item Under the type I Seesaw mechanism~\cite{Minkowski:1977sc,Yanagida:1980xy,Gell-Mann:1979vob,Mohapatra:1980yp}, left-handed neutrinos mix with heavy right-handed neutrinos, which gives heavy and measured light Majorana neutrinos. Mixing angles receive contributions from Yukawa couplings and the heavy right-handed neutrino mass matrix, which is responsible for the strong mixing of PMNS. 
	
\item Type II Seesaw~\cite{Mohapatra:1980yp,Magg:1980ut,Schechter:1980gr,Wetterich:1981bx,Lazarides:1980nt} contains heavy $SU(2)$-triplet scalars, which after being integrated out would give rise to an effective dimension-five operator and generate a small Majorana neutrino mass through the Higgs mechanism. The mixing angle is determined by Yukawa couplings between this scalar field and the lepton doublet. This indicates $\theta_{\rm p}$ should have similar divergent behavior as $\theta_{\rm c}$ since they are both originated from Yukawa couplings. This is not favored by the measured data.
	
\item Type III seesaw mechanism~\cite{Foot:1988aq,Ma:1998dn,Ma:2002pf} uses $SU(2)$-triplet fermions to suppress the measured neutrino mass. Mixing angles in this scenario are contributed from Yukawa couplings and introduced heavy mass for those fermions. The situation is similar to type I.

\end{enumerate}
	
	The study of leptogenesis is difficult due to the large number of introduced parameters and usually, those heavy parameters have nothing to do with the accessible light sector. They shall be treated as independent parameters with their own intrinsic statistical distribution in the landscape. Dynamics like the seesaw mechanism mix them with the light sector, which opens a window to probe the heavy sector. From now on, we focus on the type I seesaw mechanism and discuss the mixing feature. PMNS and CKM matrices are similar, the connection between them will be shown below.

	\subsection{CKM and PMNS Mixing Angles}
	\label{subsec:review}
	
	The quark and lepton masses in SM are generated from the Higgs mechanism through Yukawa couplings between them and Higgs doublet. The mixing between generations is intrinsic. Yukawa couplings form a three-by-three complex matrix, which could be decomposed by two unitary matrices. In the quark sector, we have $Q_{\rm L}^{i}$, $u_{\rm R}^{i}$ and $d_{\rm R}^{i}$, where $i=1,2,3$ is generation index. Choose the proper base by making the field redefinition and one could get the simplified lagrangian as
	\begin{align}
		-\mathcal{L}_{\rm q}&= Y_{\rm d}^{(k)}\bar{Q}_{\rm L}^k\cdot H d_{\rm R}^k+y_{\rm u}^{ij}\epsilon_{ab}\left(\bar{Q}_{\rm L}^{i}\right)_a H_b^\dagger u_{\rm R}^{j}+ {\rm h.c.}\;, 
		\label{eq:Yukawa_q}
	\end{align}
where $\epsilon_{12}=-\epsilon_{21}=+1$ and $a,b$ are the $SU(2)$ indices. The summation is indicated in repeated indices. Kinetic terms are neglected here. $H=(h^{-},h^0)^\top$ is the Higgs doublet and $\langle H\rangle =(0,v)^\top$, where $v$ is the Higgs vacuum expectation value. Here we originally have two Yukawa matrices and they could be decomposed totally by four unitary matrices. The field redefinition could eliminate three of them, leaving only one unitary matrix, which leads to one coupling matrix un-diagonalized,
	\begin{equation}
		y_{\rm u}=V_{\rm c}^\dagger Y_{\rm u}^{}\;,
		\label{eq:yu}
	\end{equation}
where $Y_{\rm u,d}$ are diagonal eigenvalue matrices and $V_{\rm c}$ is the CKM matrix. Further phase rotation shall reduce $5$ real DOF out of $V_{\rm c}$, leaving only three real mixing angles and one CP-violating Dirac $\delta_{\rm c}$. The general unitary three-by-three matrix with four real DOF could be expressed in the SP,
		\begin{equation}
			U_{\rm SP}\left(\theta_{ij},\delta_{13}\right)=
			\begin{pmatrix}
				c_{13}c_{12} & c_{13}s_{12} & s_{13}e^{-i\delta_{13}}\\
				-c_{23}s_{12}-s_{23}s_{13}c_{12}e^{i\delta_{13}} & c_{23} c_{12}-s_{23}s_{13}s_{12} e^{i\delta_{13}}  & s_{23}c_{13} \\
				s_{23}s_{12}-c_{23}s_{13}c_{12}e^{i\delta_{13}} & -s_{23}c_{12}-c_{23}s_{13}s_{12} e^{i\delta_{13}} & c_{23} c_{13} 
			\end{pmatrix}\;,
			\label{eq:general_ckm}
		\end{equation}
where $c_{ij}=\cos{\theta_{ij}}$ and $s_{ij}=\sin{\theta_{ij}}$. The mixing angles are defined within $0<\theta_{ij}<\pi/2$ and phase $0<\delta_{13}<2\pi$. Here we denote CKM matrix as
	\begin{equation}
		V_{\rm c}=U_{\rm SP}\left(\theta_{\rm c}^{12},\theta_{\rm c}^{23},\theta_{\rm c}^{13},\delta_{\rm c}^{}\right)\;.
	\end{equation}
Quark masses are generated from $Y_{\rm u,d}$ after Higgs acquires VEV, to which Weibull distribution provides an excellent fit and divergent power is $k_{\rm q}=0.269$~\cite{Andriolo:2019gcb}. Charged leptons are associated with neutrinos. A Weibull distribution with $k_{\ell}=0.269$ also provides a good fit to charged lepton masses with a scale different from that of quarks due to QCD.

	The neutrino is massless in SM. However, oscillation indicates that they have a tiny mass difference. Since neutrinos carry no charge, they may be the Majorana fermions.
	
	If neutrino only has Dirac mass like quark and charged lepton, then the mass term is essentially in the same form as quark, and the mixing matrix is the same as Eq.~\eqref{eq:general_ckm}. Meanwhile, neutrino masses should be sampled from the same distribution as that of charged lepton, which means that $P(m_\nu\to 0)\propto m_{\nu}^{-1+0.269}$ and they share the same overall scale. However, the current measurement of neutrino mass differences indicates otherwise.
	
	If including Majorana mass contribution, the effective neutrinos are Majorana fermions and the mass term is
	\begin{equation}
		-\mathcal{L}_{\nu,{\rm mass}}=\frac{1}{2}m^{ij}\bar{\nu}^i\nu^j+{\rm h.c.}\;,
		\label{eq:eff_nu}
	\end{equation}
where symmetric $m=m^\top$ could be decomposed by PMNS matrix as
	\begin{align}
		m&=U_{\rm p}^\top m^{\rm D} U_{\rm p}^{}\nonumber\\
		V_{\rm p}&=
		\Phi\left(\frac{\alpha_{\rm M}}{2},\frac{\beta_{\rm M}}{2},0\right)
		U_{\rm SP}\left(\theta_{\rm p}^{12},\theta_{\rm p}^{23},\theta_{\rm p}^{13},\delta_{\rm p}^{}\right)
		\;,
		\label{eq:pmns}
	\end{align}
where $\Phi(\phi_1,\phi_2,\phi_3)={\rm diag}(e^{i\phi_1},e^{i\phi_2},e^{i\phi_3})$ is the diagonal pure phase matrix and $m^{\rm D}={\rm diag}(m_1,m_2,m_3)$ is the physical neutrino mass matrix. Simplest seesaw mechanism gives roughly $m_{\nu}\sim D^2/M$, where $D$ is the Dirac mass that shared the same origin as charged lepton and $M$ is the Majorana mass contribution; so $P(m_\nu\to 0)\propto m_{\nu}^{-1+0.135}$, where the divergent power is half of that of Dirac neutrino mass~\cite{Andriolo:2019gcb}.

 Here $0<\alpha_{\rm M},\beta_{\rm M}<2\pi$ are Majorana phases which is the main differences between forms of $V_{\rm p}$ and $V_{\rm c}$. Experimentally measured values are shown in table~\ref{tab:data}. PMNS values are the results of three-$\nu$ oscillation analysis of the existing experimental data with Normal Hierarchy ($\mathcal{NH}$) and Inverted Hierarchy ($\mathcal{IH}$)~\cite{ParticleDataGroup:2020ssz}. Majorana phases $(\alpha_{\rm M},\beta_{\rm M})$ could not be measured through oscillation. Clearly, mixing angles do not violate CP, which are different from phases. There are totally three mixing angles for both $V_{\rm c}$ and $V_{\rm p}$, and that makes it just enough to discuss statistics. Majorana and Dirac phases have different physical origins. The numbers of phases are not enough to discuss the statistical property. We concentrate on mixing angles.
 
For simplicity, the error bars in table~\ref{tab:data} are treated as higher and lower limits for data points, from which we do the Monte-Carlo simulation to generate error bars for the following parameters.
	
	 From the numbers in table~\ref{tab:data}, one could see that CKM mixing angles favor small values while that of PMNS matrix favor relatively large values. Usually, we describe the CKM matrix as weak mixing between quarks and PMNS as relatively strong mixing between neutrinos. Different statistical pattern leads us to assume that these mixing angles are sampled from two different statistical distributions $P(\theta_{\rm c})$ and $P(\theta_{\rm p})$. Note that the distribution of mixing angles could be different under a different parametrization. Our goal here is to explore the heavy neutrino mass distribution, which is independent of the parametrization scheme. The convergence of different schemes will be discussed later in section~\ref{sec:diffp}. Here we choose a specific parametrization and proceed. By adopting the SP, one could quantitatively compare the mixing feature of CKM and PMNS matrices. The strength of mixing could be quantitatively described by the divergent power of distribution as in Eq.~\eqref{eq:qk}. 

\begin{itemize}	
\item When mixing angles $\theta_{ij}$ are localized around zero ($P(\theta)$ diverges at $\theta=0$), correspondingly $k<1$, the mixing between those states are small (the probability of transition is small as it is proportional to mixing angles). Also, smaller $k$ implies severer divergence in $P(\theta)$, which means weaker mixing.
	
\item $k>1$ means that distribution of $\theta_{ij}$ is not localized at zero ($P(\theta)$ is smooth at $\theta=0$), leading to a relatively strong mixing of states. Without non-trivial distribution introduced, larger $k$ implies stronger mixing.
\end{itemize}
	
\begin{table}[ph]
\tbl{Best-fit parameters transformed Beta Distribution.}
{\begin{tabular}{@{}cccc@{}}
\toprule
Shape Parameters&  & $a$ & $b$\\
\colrule
CKM & & $0.578^{+0.0043}_{-0.0033}$ & $9.55^{+0.087}_{-0.051}$ \\
\colrule
PMNS  & $\mathcal{NH}$ & $2.12^{+0.052}_{-0.043}$ & $4.80^{+0.22}_{-0.11}$ \\
	  & $\mathcal{IH}$ & $1.80^{+0.037}_{-0.030}$ & $3.63^{+0.17}_{-0.10}$ \\
\botrule
\end{tabular}\label{tab:best_fit}}
\end{table}	
	
	To extract the divergent power of intrinsic $P(\theta_{\rm c})$ and $P(\theta_{\rm p})$ from experimentally measured values, we use the \emph{transformed Beta distribution} with two shape parameters to fit the data, $P(\theta)\approx B(\theta;a,b)$, with
	\begin{align}
		B(\theta;a,b)&=\left(\frac{2}{\pi}\right)^{a+b-1}\frac{\Gamma(a+b)}{\Gamma(a)\Gamma(b)}\theta^{a-1}\left(\frac{\pi}{2}-\theta\right)^{b-1}\nonumber\\
		&\overset{\theta\to 0}{\sim} \mathcal{N} \theta^{-1+a} + \cdots\;, 
		\label{eq:beta_dist}
	\end{align}
where $\Gamma(z)$ is the gamma function and $(a,b)$ are positive real numbers. It is normalized on $\theta\in[0,\pi/2]$. $a<1$ implies distribution diverges at $\theta=0$ and $a\geq 1$ indicates smooth distribution at the origin. Parameter $b$ controls the behavior at the end point $\theta=\pi/2$ in the same way as $a$. The divergent power of $B(\theta;a,b)$ at the origin is determined by parameter $a$. There are three measured angles and two parameters in Beta distribution, one could always provide a good fit to extract information of intrinsic distribution. The best-fit value of $(a,b)$ for CKM and PMNS mixing angles are shown in table~\ref{tab:best_fit}. Here we simply take best-fit parameter $a$ as an approximation to the intrinsic divergent power of $P(\theta_{\rm c})$ and $P(\theta_{\rm p})$. Therefore, central values are
	\begin{equation}
		k_{\rm c}\approx 0.578\;,\quad k_{\rm p}\approx 
		\begin{cases}
			2.12 & \mathcal{NH}\\
			1.80 & \mathcal{IH}\\
		\end{cases}\;.
		\label{eq:kckp}
	\end{equation}

	Here $k_{\rm c}<1$ shows a diverging distribution of $\theta_{\rm c}$ at origin. If we take peaking behavior of distribution as an intrinsic property of SM, this is natural. However, $k_{\rm p}>1$ indicates a clear different in origin between CKM and PMNS mixing angles, even though they are expressed in the similar form. Type I and III seesaw introduce extra sector to lower the neutrino mass. How $\theta_{\rm p}$ is affected by heavy sector depends on specific model set-up. We will discuss type I due to the fact the link between light and heavy sector is clearer than that in type III and it is widely discussed in the context of leptogenesis.

	\section{An $O(2)$ type I Seesaw Toy Model}
	\label{sec:o2}
	
	We take a detailed look at an $O(2)$ type I seesaw toy model. The basic idea of the seesaw mechanism is using a heavy sector which is integrated out in low energy effective theory to suppress the effective neutrino mass. Type I seesaw adds heavy right-handed neutrinos to SM and it naturally fits in GUT theory. In common leptogenesis scenarios, heavy Majorana neutrinos are needed to generate the violation of lepton number. Here this simple $O(2)$ type I seesaw model would illustrate the effects on mixing angles in PMNS from the heavy sector.

	Similar as Eq.~\eqref{eq:Yukawa_q}, the general Lagrangian of lepton Yukawa sector in the presence of right hands neutrino $N$ is given by 
	\begin{equation}
		\label{eq:LagrangianOringin}
		-\mathcal{L}_{\ell}=Y_{e}^{(k)}\bar{\ell}_{\rm L}^{k}\cdot H e_{\rm R}^{k}+\tilde{\lambda}^{ij} \epsilon_{ab} (\bar{\ell}_{\rm L}^{i})_a^{} H_b^\dagger N^j+\frac{1}{2} \mathcal{M}^{ij} \bar{N}^i N^j+{\rm h.c.}\;,
	\end{equation}
where ${\ell}_{\rm L}^{i}$ refers to $SU(2)_{\rm L}$ lepton doublets and $e_{\rm R}^j$ is the singlet, and Majorana mass for $N$ is symmetric, $\mathcal{M}^{ij}=\mathcal{M}^{ji}$. Above Lagrangian has already chosen a base where down-part of doublet $\ell_{\rm L}^{i}$ already diagonalized like in quark sector Eq.~\eqref{eq:Yukawa_q}, which makes the coupling in the same form as Eq.~\eqref{eq:yu},
	\begin{equation}
		\tilde{\lambda}={V_{\rm c}'}^\dagger\lambda^{\rm D}_{}\;,
	\end{equation}
	where $\lambda^{\rm D}$ is the diagonal eigenvalue matrix. Without the Majorana mass term of right-handed neutrino, $\mathcal{M}^{ij} \to 0$, above Lagrangian is in the exact same form as that of quarks~\eqref{eq:Yukawa_q}. Thus, $P(\lambda^{\rm D}_i\to 0)\propto (\lambda^{\rm D}_i)^{-1+0.269}$, where the divergent power is the same as that of quark and lepton mass, inherited from Yukawa coupling. $V_{\rm c}'$ is the mixing matrix which has the same nature as CKM and  parameters in it should be sampled from the same distribution as that in CKM. We use subscript `c' to denote parameters of CKM type.

	Here the toy model only involve two generations and real couplings, which means that the mixing matrix is described as a two-by-two rotational matrix, $RR^\top=1$,
	\begin{equation}
		R(\theta)=
		\begin{pmatrix}
			\cos{\theta} & -\sin{\theta}\\
			\sin{\theta} & \cos{\theta}
		\end{pmatrix}\;.
	\end{equation}
The symmetric Majorana mass matrix could be decomposed by another mixing angles as
	\begin{equation}
		\mathcal{M}=U_{\rm N}^\top 
		\begin{pmatrix}
			M_1 & 0\\
			0 & M_2
		\end{pmatrix} U_{\rm N}^{}\;,
	\end{equation}
where $U_{\rm x}=R(\theta_{\rm x})$ is the abbreviation for any label `x'. Further field redefinition $N\to U_{\rm N}^\top N$ would diagonalize right-handed neutrino and makes the $\tilde{\lambda}$ receive rotation from the right,
	\begin{equation}
		\tilde{\lambda}\to\lambda=U_{\rm c}^\top\lambda^{\rm D} U_{\rm N}^{}\;,
		\label{eq:lambda}
	\end{equation}
where $U_{\rm c}$ is real 2D version of $V_{\rm c}'$. The effective neutrino mass matrix is obtained through seesaw mechanism,
	\begin{equation}
		m=\lambda
		\begin{pmatrix}
			\cfrac{1}{M_1} & 0\\
			0 & \cfrac{1}{M_2}
		\end{pmatrix}
		\lambda^\top\;,
	\end{equation}
which is also symmetric as expected and could be decomposed by a PMNS angle $\theta_{\rm p}$ like \eqref{eq:pmns}. Without complex phases, PMNS and CKM matrix are in the exactly same form but with their mixing angles from different origin. By inserting Eq.~\eqref{eq:lambda} and comparing it with $m=U_{\rm p}^\top m^{\rm D} U_{\rm p}^{}$, one could conclude that
	\begin{equation}
		U_{\rm p}=U_{\beta}U_{\rm c}\;,\quad\theta_{\rm p}=\theta_{\rm c}+\beta\;,
		\label{eq:upbc}
	\end{equation} 
where $U_\beta$ is the effective contribution that is determined by $\lambda^{\rm D}$, $M_{1,2}$ and $U_{\rm N}$, explicitly
	\begin{equation}
		m^{\rm D}
		=
		U_\beta^{}
		\lambda^{\rm D}
		U_{\rm N}^{}
		\begin{pmatrix}
			\cfrac{1}{M_1} & 0\\
			0 & \cfrac{1}{M_2}
		\end{pmatrix}
		U_{\rm N}^\top
		\lambda^{\rm D} U_\beta^\top
		\;.
	\end{equation}
Here $\lambda^{\rm D}={\rm diag}(\lambda_1,\lambda_2)$ and $m^{\rm D}={\rm diag}(m_-,m_+)$. Vanishing off-diagonal element of $m^{\rm D}$ gives us the relation
	\begin{equation}
		\tan(2\beta)\equiv f\left(\eta,\xi,\theta_{\rm N}\right)=\frac{2 \eta  (\xi -1) \tan{\theta_{\rm N}}}{\left(\eta ^2-\xi\right)+\left(\eta ^2 \xi -1\right) \tan^2{\theta_{\rm N}}}\;,
		\label{eq:f_b}
	\end{equation}
where $\eta=\lambda_1/\lambda_2$ and $\xi=M_1/M_2$. Here we demand that $0<\eta<1$ and allow $\xi$ to be smaller or bigger than one, since the mass hierarchy is either the same or the opposite for light and heavy neutrino. The eigenvalues of light neutrino are also given by
	\begin{align}
		m_\pm&=\frac{\lambda_2^2}{4M_1}\left(w\pm \sqrt{w-16\eta^2\xi}\right)\label{eq:mpm}\\
		w&=\left(\eta^2+1\right) \left(\xi +1\right)-\left(\eta^2-1\right) \left(\xi -1\right)\cos{(2\theta_{\rm N})}\;.\nonumber
	\end{align}
Notice that there exist some constraints on the allowed value of $\eta$, $\xi$ and $\theta_{\rm N}$ due to the fact that eigenvalues should be semi-positive, $m_\pm \geq 0$. Including the complex phase would save the day and phases exist in real life. This toy model serves as a leading-order approximation to explore the relations between the heavy and the light sector. Thus, we focus on the relation given by function $f$ in Eq.~\eqref{eq:f_b} and allow parameters to vary in the whole parameter space, neglecting those constraints.
	
	This $O(2)$ model shows explicitly that $\theta_{\rm p}$ receives contribution from heavy sector as Eq.~\eqref{eq:upbc} and Eq.~\eqref{eq:f_b}. The extension to $U(3)$ real case is complicated. However, for our purpose, this toy model is sufficient.

	\section{Divergent Powers}\label{sec:k}
	
	\subsection{Implication of Relations between Parameters}
	
	For parameters that are generated through algebraic relations from their `atom'-parameters, which are sampled from their own distribution, the divergent powers of them also satisfy specific relations. Here we use the relation obtained from $O(2)$ toy model in the last section to make an estimate.
	
	The divergent power for summation of independent variates are the sum of each variate's divergent power. Some examples are shown in \ref{appendix:stat}. Since $\theta_{\rm p}$ is the sum of $\theta_{\rm c}$ and intermediate angle $\beta$ in this $O(2)$ model,
	\begin{equation}
		k_{\rm p}=k_{\rm c}+k_\beta\;,
		\label{eq:kpcb}
	\end{equation}
where $\beta$ is the composite of $\xi$, $\eta$ and $\theta_{\rm N}$ as in Eq.~\eqref{eq:f_b}. The contribution from $\beta$ changes the divergent power of $\theta_{\rm p}$ and makes it not peaking at the origin. For the three-generation case, above relation \eqref{eq:kpcb} is not exact. From the MC simulation in \ref{appendix:mc_sim}, one could conclude that the resulting divergent power for $\theta_{\rm p}$ is dominated by the sum of $k_{\rm c}$ and $k_\beta$ in the $SO(3)$ case. Thus, Eq.~\eqref{eq:kpcb} is a good approximation.
	
	The distribution $P(\beta)$ is obtained from $P(\xi)$, $P(\eta)$ and $P(\theta_{\rm N})$ through Eq.~\eqref{eq:f_b}, and it could be calculated using Eq.~\eqref{eq:py},
		\begin{align}
			P(\beta)&=\int{\rm d} \xi \,{\rm d} \eta \,{\rm d} \theta_{\rm N}\,P(\xi) P(\eta)P(\theta_{\rm N})\,\delta\left(\beta-\frac{1}{2}\tan^{-1}f\right) \nonumber\\
			&\approx \mathcal{N}\left(1+\tan^2{2\beta}\right) \int_{\mathcal{S}} {\rm d} \xi\, {\rm d} \eta\,\xi^{-1+k_\xi} \eta^{-1+k_\eta}\nonumber\\
			&\qquad \qquad\times \int_0^{\pi/2} {\rm d} \theta_{\rm N}^{}\,  \theta_{\rm N}^{-1+k_{\rm N}}\frac{\delta\left(\theta_{\rm N}-\theta_0\right)}{\left|f'(\theta_0)\right|}\;,
			\label{eq:p_beta}
		\end{align}
where $\theta_0$ satisfies $\tan{2\beta}=f(\theta_0)$ and normalization $\mathcal{N}$ is irrelevant here. $f'$ denotes derivative of $f$ with respective to $\theta_{\rm N}$. Parameters $\xi$, $\eta$ and $\theta_{\rm N}$ are directly generated from flux in this model; so they are independently sampled from their own distributions. The peaking property of their distribution is generic. (In fact, divergent power of $\eta$ is determined by that of $\lambda^{\rm D}_i$, which is the part that assumed to be exactly the same as quark sector. So $k_\eta=k_{\rm q}=0.269$.) Therefore, $P(\xi)$, $P(\eta)$ and $P(\theta_{\rm N})$ are approximated using \eqref{eq:qk} in Eq.~\eqref{eq:p_beta}. The detailed form of $P(\beta)$ may be complicated. Since we only need divergent power of $\beta$, the analysis could be carried out straight forwardly.

	Consider $\xi$ and $\eta$ as parameters and then $\beta$ is a function of $\theta_{\rm N}\in [0,\pi/2]$ only. Trigonometric function $\tan{2\beta}$ has period of $\pi$. Thus, $\beta$ is confined at $\beta\in [0, \pi/2]$. In order to let $\beta$ cover the whole range, there must exist a point $\theta_{\rm crit}$ such that
	\begin{align}
		\lim_{\epsilon\to 0}f\left(\theta_{\rm crit}-\epsilon\right)&=+\infty \nonumber\\
		\lim_{\epsilon\to 0}f\left(\theta_{\rm crit}+\epsilon\right)&=-\infty \;.\label{eq:theta_crit}
	\end{align}
If the function $f(\theta_{\rm N})$ does not blow up within $0\leq\theta\leq \pi/2$, the possible value of $\beta$ is even limited smaller and may not get $\beta=0$. It depends on the parameter region of $(\eta,\xi)$. Here we divide the parameter space into four regions $\mathcal{S}=\mathcal{S}_1+\mathcal{S}_2+\mathcal{S}_3+\mathcal{S}_4$, which are
	\begin{align}
		\mathcal{S}_1&=\left\{(\eta,\xi)\;\bigg|\;\xi>1\;,\;\; 0<\eta^2<\frac{1}{\xi} \right\}\;\nonumber\\
		\mathcal{S}_2&=\left\{(\eta,\xi)\;\bigg|\;\xi>1\;,\;\; \frac{1}{\xi}<\eta^2<1 \right\}\;\nonumber\\
		\mathcal{S}_3&=\left\{(\eta,\xi)\;\bigg|\;\xi<1\;,\;\; 0<\eta^2<\xi \right\}\;\nonumber\\
		\mathcal{S}_4&=\left\{(\eta,\xi)\;\bigg|\;\xi<1\;,\;\; \xi<\eta^2<1 \right\}\;,
		\label{eq:subspace}
	\end{align}
where we have already let $0<\eta<1$ by definition. $\xi>1$ corresponds to the situation that $M_1>M_2$, which is the heaviest right-handed (does not carry $SU(2)$ charge) neutrino corresponds to lightest left-handed (in $SU(2)$ doublet) neutrino.
	
	\begin{figure}
	\centering
		\includegraphics[width=8.6cm]{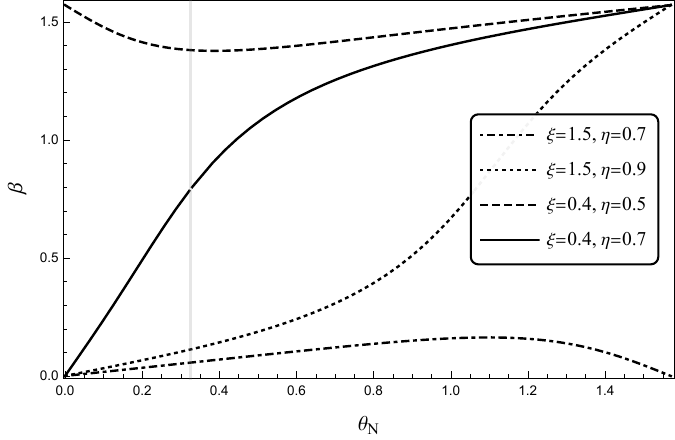}
		\caption{The plot of $\beta$ as function of $\theta_{\rm N}$ within four different parameter subspace. Dot-dashed, dotted, dashed and solid curves correspond to $\mathcal{S}_1$, $\mathcal{S}_2$, $\mathcal{S}_3$ and $\mathcal{S}_4$ subspace defined in Eq.~\eqref{eq:subspace}. Vertical gray line is the $\theta_{\rm crit}$ in Eq.~\eqref{eq:theta_crit} for solid curve.}
		\label{fig:beta_f}
	\end{figure}	
	
	Four parameter-subspaces give different types of $\beta(\theta_{\rm N})$ curves, which are plotted in figure~\ref{fig:beta_f}. Here we discuss these four scenarios.
	
\begin{itemize}

\item Curves $\beta(\theta_{\rm N})$ in $\mathcal{S}_1$ starts and ends at zero for $\theta_{\rm N}=0$ and $\theta_{\rm N}=\pi/2$ respectively. $\eta<1/\sqrt{\xi}$ and smaller $\eta$ corresponds to  a $\beta(\theta_{\rm N})$ curve closer to the $\beta=0$ line. Since  $P(\eta)$ is peaking at zero ($k_\eta \approx 0.269<1$), small $\beta$ is guaranteed.  In this region, value of $\xi$ close to one also leads to vanishing $\beta$. This indicates that in this region of parameter space, it is easily to obtain a small value of $\beta$ that is close to zero and the divergent power $k_\beta$ should in general small than 1.
	
\item In $\mathcal{S}_2$, $\beta(\theta_{\rm N})$ starts at origin and ends in $\beta=\pi/2$. Here $1/\sqrt{\xi}<\eta<1$, so the value of $\eta$ does not cause the vanishing $\beta$. Also, $P(\xi)$ is dominated at values close to one, which gives vanishing $\beta$. Thus, $k_\beta$ is mainly determined by $P(\xi\to 1)$ and $k_{\rm N}$.
	
\item Subspace $\mathcal{S}_3$ does not contain vanishing $\beta$, which shall be simply neglected.
	
\item $\mathcal{S}_4$ gives the desired scenario. Curve $\beta(\theta_{\rm N})$ starts at origin and ends at $\beta=\pi/2$. Vanishing $\theta_{\rm N}$ gives vanishing $\beta$. Parameter $\eta>\sqrt{\xi}$ is bounded from below, which makes $k_\eta$ do not contribute to $k_\beta$. $\xi\to 1$ could acquire small $\beta$. However, unless $k_\xi\gg 1$ or some non-trivial form is introduced  , $P(\xi)$ within this region could be treated simply as smooth (no divergence or severe peak). Therefore, $k_\beta$ is dominated by $k_{\rm N}$ and it could be larger than one.
	
\end{itemize}

If one considers the situation $\xi>1$, the peaking distribution of $\beta$ is guaranteed due to the inclusion of $\mathcal{S}_1$. To explain the measured relatively large $\theta_{\rm p}$, we need bigger $\beta$ (or $P(\beta)$ with big enough $k_\beta$) to diminish the peaking behavior of $\theta_{\rm p}$. Once this scenario is allowed, it would require a high-level fine-tune to achieve $k_{\rm p}>1$. Therefore, we discard this case. Physically, one could conclude that the mass hierarchy of heavy neutrinos should naturally be the same as light neutrinos and charged leptons.
	
	From now on, we let $\xi<1$ and focus on small $\beta$ region. In order to get $\beta=0$, we only need to focus on $\theta_{\rm N}=0$ in $\mathcal{S}_{4}$. The function $f(\theta_{\rm N})$ could be expanded for simplicity, as
	\begin{equation}
		f(\theta_{\rm N})\approx \frac{\eta\left(1-\xi\right)}{\eta^2-\xi}\theta_{\rm N}+\mathcal{O}\left(\theta_{\rm N}^2\right)\;.
	\end{equation}
Take the leading order and the integral in Eq.~\eqref{eq:p_beta} could be evaluated as
	\begin{align}
		P(\beta)&\overset{ \beta\to 0 }{\sim}\mathcal{N} I \beta^{-1+k_{\rm N}}+\cdots\\
		I&~=\int_{0}^{1}{\rm d}\xi \int_{\sqrt{\xi}}^{1}{\rm d} \eta \;\xi^{-1+k_\xi}\eta^{-1+k_\eta}\left[\frac{\eta^2-\xi}{\eta(1-\xi)}\right]^{k_{\rm N}}\;.\nonumber
	\end{align}
The integral $I$ contains no $\beta$ and serves simply as a pre-factor. Therefore, together with Eq.~\eqref{eq:kpcb}, one could conclude that
	\begin{equation}
		k_{\rm N}=k_{\beta}=k_{\rm p}-k_{\rm c}\;.
		\label{eq:kk}
	\end{equation}
	Distribution of heavy neutrino mixing angles $P(\theta_{\rm N})$, together with that of Yukawa matrix (CKM-like angles and eigenvalues), form the measured PMNS matrix, which causing the distinct distribution pattern between PMNS and CKM mixing angles. The interference information is indicated by divergent power.

\subsection{Probing The Heavy Neutrino}
	
Since mixing angles $\theta_{\rm N}$ and eigenvalues $M_i$ are decomposed from heavy neutrino mass matrix in Eq.~\eqref{eq:LagrangianOringin}, the divergent power of their distributions should obey some numerical relations. Here we propose a linear relation, which provides an excellent fit to the simulation, to do estimation and more analysis could be found in \ref{appendix:mc_sim}. Numerical fit tells us
\begin{equation}
		k_{\rm M}\approx 0.75 k_{\rm N}+0.11\;,
		\label{eq:linear_k}
\end{equation}
where $k_{\rm M}$ is the divergent power for heavy neutrino mass distribution $P(M)$. Using the values in Eq.~\eqref{eq:kckp} and relation Eq.~\eqref{eq:kk}, one could obtains (central values)
	\begin{equation}
		k_{\rm M} \approx
		\begin{cases}
		1.26 & \mathcal{NH}\\
		1.02 & \mathcal{IH}
		\end{cases} \;,
	\end{equation}
There is a difference between $\mathcal{NH}$ and $\mathcal{IH}$. Despite that, $k_{\rm M}\gtrsim 1$ shows that the distribution of $M$ is not peaking at the origin and indicates a degenerate mass pattern. This is important and will be demonstrated below.
		
Despite the overall scale of $M$, the patten of heavy neutrino mass $M_i$s is mainly dictated by divergent power. Since $k_{\rm M}\gtrsim 1$, $P(M)$ is smoothly distributed around the origin. Mass is an unlimited parameter; so there must exist a tail suppression for $P(M)$. As shown in figure~\ref{fig:main}, solid-line around $M=0$ is obtained from $k_{\rm M}\approx 1.3$ and solid-line at large $M$ is the suppression for the sake of normalization of $P(M)$. Some non-trivial forms could exist in the middle, where the region is labeled gray. Naturally, one would expect a bump in the gray region, which indicates the localization of heavy neutrino masses.
	
	Since the mass parameter is unbounded and Weibull distribution provides a good fit to quark and lepton mass distribution, it is reasonable to do the same here. We approximate the distribution of $M_i$ as $P(M)\approx \tilde{P}(M; k,l)$ to quantitatively demonstrate the localization of $P(M)$. Here $k$ encodes the divergent power and $l$ is the overall scale of this heavy sector, which is unknown. Fortunately, the scale is irrelevant when discussing the relative ratio between masses.
	
\begin{figure}
		\centering
		\includegraphics[width=8.6cm]{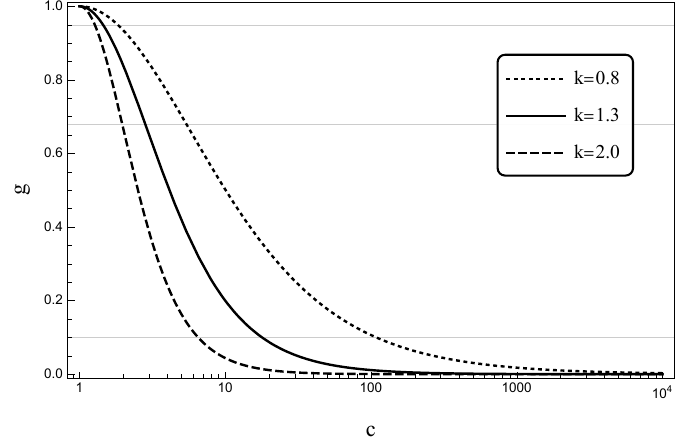}
		\caption{Plot of confidence probability $g$ of $M_3\geq c M_1$ with different divergent power $k$. The plot shows that the confidence probability is small for $c$ having large value, especially for large $k_{\rm M}$. Three gray horizontal lines correspond to probability $95\%$, $68\%$ and $10\%$, from top to bottom.}
		\label{ratiofig}
	\end{figure}

If our Universe has only three heavy neutrinos in total, the joint probability of three mass distribution is described by direct product of these three distributions,
	\begin{align}
		P_{123}&=P(M_1)P(M_2)P(M_3)\\
		P(M_i)&\approx \tilde{P}=\frac{k}{l}\left(\frac{M_i}{l}\right)^{-1+k}\exp\left[-\left(\frac{M_i}{l}\right)^{k}\right]\;.\nonumber
	\end{align}
Define a function $g(c)$ which describes the probability that
	\begin{equation}
		M_3\geq c M_1\;,
	\end{equation}
where $c\geq1$ and by default $M_3>M_2>M_1$. This could be directly calculated by integration
	\begin{align}
		g(c)&= 6\int_0^\infty{\rm{d}}M_1\int_{M_1}^{M_3}{\rm{d}}M_2\int_{cM_1}^\infty{\rm{d}}M_3 P_{123}\nonumber\\
		&\approx\frac{9c^{k}}{(2+c^{k})(1+2c^{k})}\;,
		\label{eq:g}
	\end{align}
where $6$ before the integral is the normalization factor such that $g(1)=1$. Figure~\ref{ratiofig} shows the plot of $g(c)$ with different $k$. We can see when $k=1.3$, the probability of $M_3\geq 10 M_1$ is lower than $20\%$ and the probability of $M_3\geq 100 M_1$ is lower than $2\%$. Higher $k$ indicates strong localization of $g$ on smaller $c$. That is, the pattern of heavy neutrino mass is expected to be degenerate for $k_{\rm M}\gtrsim 1$. Some numerical results are exhibited in table~\ref{tab:example}.

\begin{table}[ph]
\tbl{Examples for different probability of heavy sector mass ratio using Weibull distribution.}
{\begin{tabular}{@{}ccc@{}}
\toprule
$k$ & Mass relation & Probability($\%$) \\
\colrule
$0.8$ & $M_3\geq 1.83M_1$  & $95$\\
	  & $M_3\geq 5.41 M_1$ & $68$\\
	  & $M_3\geq 108 M_1$  & $10$\\
\colrule
$1.3$ & $M_3\geq 1.45 M_1$ & $95$\\
	  & $M_3\geq 2.83 M_1$ & $68$\\
	  & $M_3\geq 17.9 M_1$  & $10$\\
\colrule
$2.0$ & $M_3\geq 1.27M_1$ & $95$\\
	  & $M_3\geq 1.96M_1$  & $68$\\
	  & $M_3\geq 6.52M_1$  & $10$\\
\botrule
\end{tabular}\label{tab:example}}
\end{table}

\section{Different parametrization}
\label{sec:diffp}

The specific parameters for CKM and PMNS mixing angles and phases are dependent on parametrization. The physics here is the distinct `mixing strength' of these two matrices. CKM is weak-mixing and PMNS is relatively strong-mixed. To compare these two matrices, we choose the standard parametrization (SP) for both of them and investigate the divergent powers of their distribution in previous sections. The resultant heavy neutrino masses distribution has the property $k_{\rm M}\gtrsim 1$, which indicates a degenerate mass pattern. Here we would like to show that the conclusion for $k_{\rm M}$ still hold, if we starting from a different parametrization. We seek the parametrization that reflects the unitarity, which apparently Wolfenstein parametrization does not apply here. Here we take KM parametrization as a comparison.

KM parametriztion (KM) used in CKM matrix is given by unitary matrix below, compared to Eq.~\eqref{eq:general_ckm},
\begin{equation}
U_{\rm KM}\left(\theta_{i},\delta\right)=
\begin{pmatrix}
c_1 && -s_1 c_3 && -s_1 s_3\\
s_1 c_2 && c_1 c_2 c_3 - s_2 s_3 e^{i\delta}   && c_1 c_2 s_3 + s_2 c_3 e^{i\delta} \\
s_1 s_2 && c_1 s_2 c_3 + c_2 s_3 e^{i\delta} && c_1 s_2 s_3 - c_2 c_3 e^{i\delta}
\end{pmatrix}\;,
\label{eq:KMP}
\end{equation}
where $c_j=\cos{\theta_j}$ and $s_j=\sin{\theta_j}$. Then for CKM and PMNS matrices, one has 
\begin{align}
V_{\rm c}&=U_{\rm KM}\left(\theta_{{\rm c},1},\theta_{{\rm c},2},\theta_{{\rm c},3},\delta_{\rm c}\right)\nonumber\\
V_{\rm p}&=\Phi\left(\frac{\alpha_{\rm M}}{2},\frac{\beta_{\rm M}}{2},0\right)U_{\rm KM}\left(\theta_{{\rm p},1},\theta_{{\rm p},2},\theta_{{\rm p},3},\delta_{\rm p}\right)\;.
\end{align}
Due to the lack of global analysis, we simply convert the data in table~\ref{tab:data} from SP to KM by proper phase rotation of original matrices from the left or right.

\begin{figure}
\centering
\includegraphics[width=8.6cm]{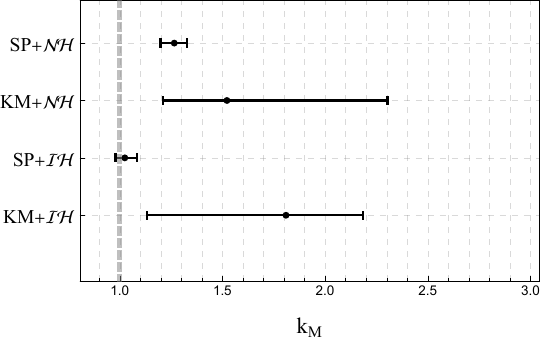}
\caption{The resultant $k_{\rm M}$ with $1\sigma$ error bar transmitted in SP and KM under $\mathcal{NH}$ and $\mathcal{IH}$.}
\label{fig:errorbar}
\end{figure}

The difference for $k_{\rm c}$ and $k_{\rm p}$ in KM and SP are obvious and they shall be compensated by different relation between $k_{\rm N}$ and $k_{\rm M}$ in different parametrization. Formally, one could conclude the following master equation,
\begin{equation}
k_{\rm M}=h\left(k_{\rm N}\right)=h\left(k_\beta\right)=h\left(k_{\rm p}-k_{\rm c}\right)\;,
\label{eq:kkkk}
\end{equation}
where the function $h$ depends on the parametrization as indicated in Eq.~\eqref{eq:kmkn}. By adopting the different form of $h$ and different fitted value of $(k_{\rm p},k_{\rm c})$ in KM and SP, we finally obtains the figure~\ref{fig:errorbar}. For other parametrizations, both the starting point (the distribution of mixing angles) and the analysis path (function $h$) are different, and they shall give the same conclusion, $k_{\rm M}\gtrsim 1$.

As shown by the figure~\ref{fig:errorbar}, all central values are greater than $1$. Error bars in KM are larger than that in SP, which is due to the sloppy direct conversion we performed here for simplicity. More careful global fit in KM would shrink error bars. The resultant $k_{\rm M}$s in $\mathcal{NH}$ essentially agree with each other, and this shows a convergence on different parametrization.

The huge error bar of the Dirac phase in PMNS, due to the difficulty in precision measurement, causes the conversion between SP and KM to be unstable, which is also responsible for the disagreement in SP+$\mathcal{IH}$ and KM+$\mathcal{IH}$ in $1\sigma$ range. Also, the Seesaw mechanism mixes Dirac and Majorana mass contribution in a way such that both of them would contribute to the light neutrino mass. According to Eq.~\eqref{eq:mpm}, $m_\nu \sim \lambda^2/4M$, which indicates that $k_\nu=k_{\rm q}/2\approx 0.135<1$ since only numerator would affect divergent power as shown in \ref{appendix:stat}. This tells us that the distribution of light neutrino mass is peaked at the origin, which indicates a similar pattern as that of leptons. Therefore, $\mathcal{IH}$ for light neutrino mass is not favored at the beginning of this analysis.

More carefully speaking, there exists a small possibility that one samples out $\mathcal{IH}$ light neutrino mass starting from a peaking distribution. To analyze the relation of divergent powers in this scenario, one has to take this conditional probability into account when deriving all the numerical relations in the previous sections, as the form of function $h$. Here we simply ignore this preliminary since $\mathcal{NH}$ is dominant when randomly sampling eigenvalues from a peaking distribution.

Overall, the core result of this work obtained from SP in previous sections, $k_{\rm M}\gtrsim 1$ (or degenerate heavy neutrino mass pattern), should be independent of parametrization.

	\section{Discussion}
	\label{sec:discussion}
	
	Some discussions are included below.
	
	\begin{itemize}
	
	\item The standard parametrization~\cite{ParticleDataGroup:2020ssz} of CKM and PMNS matrices is adopted in this work. Different parametrization would result in a different distribution for mixing angles. This causes no trouble. Choosing the parametrization is equivalent to choosing a gauge, which would lead to different descriptions of the same physical system. For our purpose, the distribution of mixing angles is a measure of strength of mixing of the original CKM or PMNS matrix. The same parametrization is adopted for both CKM and PMNS matrices in this work, which makes it reasonable to compare mixing patterns of them.
	
	\item Beta distribution $B(x;a,b)$ used in the section \ref{subsec:review} and appendix \ref{appendix:mc_sim} serve as a tool to extract the information of parameters' intrinsic distribution. The best-fit parameter $a$ is an approximation of the real divergent power $k$. With potentially more data coming in, one could expect that the information on intrinsic distribution would become more precise.
	
	\item Quark mass and lepton mass are well described by Weibull distribution. It is natural to extend the distribution to heavy neutrino masses, which is used in section \ref{sec:k}. We use Weibull distribution as a demonstration to show the impact of having a bigger than one divergent power on the statistical pattern. Without a complicated mechanism, we argue that the degenerate mass pattern of heavy neutrino is favored. Here the ``degenerate" is defined as the opposite to hierarchical ($M_3 \gg M_2 \gg M_1$). It means that heavy neutrino masses are close to each other relatively. Leptogenesis should consider all of their decay processes, instead of just considering the decay of the lightest one.
	
	\item The intrinsic distribution of the heavy neutrino sector calculated here is quite different from usual sectors such as quarks and leptons. Both $P(M)$ and $P(\theta_{\rm N})$ are believed to be not diverging at origin, which is an implication that they may not be directly generated from flux values if we take peaking behavior as a signature from string theory. Some UV structure is responsible for the appearance of the heavy neutrino sector.
	
	\item The key feature of a distribution emphasized here is divergent power at the origin. When parametrizing CKM and PMNS matrices, one usually would encounter the issue of period of trigonometric function. The periodicity would affect the allowed range for mixing angles which determines the `tail' behavior of the probability distribution. Meanwhile, the divergent power describes the `head' behavior, which encodes the `string signature'. We force the range for mixing angles in CKM and PMNS matrices to be the same for the sake of comparison.	
	
	\end{itemize}

	\section{Summary}
	\label{sec:summary}
	
	In summary, motivated by string theory, we consider an effective parameter landscape that gives birth to the statistical interpretation of couplings or parameters of low energy effective theories. This provides us with a useful tool to explore the sectors that have not been measured yet. Oscillating massive neutrino forced us to expand the Standard Model by including heavy degrees of freedom and those heavy particles may turn out to be the main players in explaining the matter-antimatter asymmetry puzzle. The PMNS matrix receives contributions from both Yukawa-like couplings and heavy Majorana masses. That makes it a portal to investigate the decoupled sector at the low energy regime. The key lies in the statistical distribution of various parameters. This work chooses divergent power of distribution at the origin as the compass, which reflects the algebraic relation according to the specific model set-up and guides us to take a glance at the unknown sea of physics beyond the Standard Model.
	
	In the minimal type I seesaw mechanism (CP-violating phases neglected), the different statistical patterns between CKM and PMNS mixing angles are expected due to the involvement of heavy Majorana neutrinos. Monte-Carlo simulations serve as thought experiments that motivate us to propose the linear relation Eq.~\eqref{eq:linear_k} (or function $h$). Simulations also show that the two-generation $O(2)$ model is a good approximation to simplify the analysis. The divergent power of heavy neutrino mass distribution is found to be larger than one, $k_{\rm M}\gtrsim 1$, which is based on the statistical pattern of measured PMNS mixing angles. Therefore, a degenerate heavy neutrino mass pattern is expected. This work suggests that leptogenesis should focus on the degenerate heavy neutrino mass scenario. Also, this work may serve as a bottom-up hint for UV model building.
	
	CP-violating phases are not considered here since there is only one in the CKM matrix and maybe two more in the PMNS matrix. Based on the current knowledge, it is difficult to speculate the distribution of CP-violating phases. It is important to include the phases due to the fact that it may give rise to some non-trivial distribution pattern of mixing angles through the seesaw mechanism, which may require further detailed study.

\section*{Acknowledgments}

We thank S. -H. Henry Tye for early involvement in this work. We thank Xuhui Jiang, Sijun Xu and Kaifeng Zheng for valuable discussions. This work is supported in part by the AOE grant AoE/P-404/18-6 issued by the Research Grants Council (RGC) of the Government of the Hong Kong SAR China. This work is also partially supported by a grant from the RGC of the Hong Kong SAR, China (Project No. 16303220).

	\appendix
	
	\section{Divergent Power of Statistical Distribution}
	\label{appendix:stat}
	
	Consider the probability distribution $P_i(x_i)$ for a positive random variate $x_i$. The no-scale assumption for flux parameter indicates that the characteristic distribution for flux parameter is either flat or peaked at the origin. It is sufficient for us to write down below distribution for the discussion in this paper.
	\begin{equation}
		P_i(x_i)=\begin{cases}
			\mathcal{N} x_i^{-1+k_i}& 0\leq x \leq \Lambda\\
			0 & {\rm otherwise}
		\end{cases} \;,
		\label{eq:appendix_px}
	\end{equation}
	where $\Lambda$ is the physical cut-off and $\mathcal{N}(k_i,\Lambda)$ is the normalization constant such that $\int P(x){\rm d}x=1$. The cut-off is usually the physical limit like $M_{\rm Pl}$. One could also use exponential factor to suppress the tail like Weibull distribution, which makes the calculation complicated. Here $0<k_i<1$ indicates a divergent distribution at the origin $x=0$ and $k_i\geq1$ for smoothly distribution at origin. There may be some tailing suppression behavior in $P(x)$ and that would change the form of peaking expression. The divergent power $k_i$ is independent of tailing behavior of $P(x)$. Here we lay out a few examples.
	
	\begin{enumerate}
	\item Suppose $y=x_1+x_2$, where $x_{1,2}$ are random variate submitted to Eq.~\eqref{eq:appendix_px}. The distribution for $y$ is calculated by
	\begin{align}
		P(y)&=\mathcal{N} \int_0^{\Lambda} x_1^{-1+k_1}x_2^{-1+k_2}\delta\left(y-x_1-x_2\right){\rm d}x_1{\rm d}x_2\nonumber\\
		&\overset{y\to 0}{\sim}\mathcal{N}\int_{0}^{y}x_1^{-1+k_1}\left(y-x_1\right)^{-1+k_2}{\rm d}x_1\nonumber\\
		&=\mathcal{N}y^{-1+k_1+k_2}\frac{\Gamma(k_1)\Gamma(k_2)}{\Gamma(k_1+k_2)}\;,
	\end{align}
	which leads to 
	\begin{equation}
		k_y=k_1+k_2\;.
	\end{equation}
	
	\item The distribution of product $z=x_1x_2$, where $x_{1,2}$ obey Eq.~\eqref{eq:appendix_px}, could also be calculated as
	\begin{align}
		P(z)&=\mathcal{N} \int_0^{\Lambda} x_1^{-1+k_1}x_2^{-1+k_2}\delta\left(z-x_1x_2\right){\rm d}x_1{\rm d}x_2\nonumber\\
		&=\mathcal{N}z^{-1+k_2}\int_{z/\Lambda}^{\Lambda} x_1^{-1+k_1-k_2}{\rm d}x_1\nonumber\\
		&\overset{z\to 0}{\propto}
		\begin{cases}
			z^{-1+k_1}\left(-\ln{z}\right) & k_1=k_2\\
			z^{-1+k_2} & k_1>k_2\\
			z^{-1+k_1} & k_1<k_2
		\end{cases}\;.
	\end{align}
	The log-divergence is weaker than the power divergence. The divergent power for production relation reads
	\begin{equation}
		k_z={\rm min}\left(k_1,k_2\right)\;.
	\end{equation}
	
	\item Statistical distribution of ratio $w=x_1/x_2$ is given by
	\begin{align}
		P(w)&=\mathcal{N}\int_0^{\Lambda}x_1^{-1+k_1} x_2^{-1+k_2}\delta\left(w-\frac{x_1}{x_2}\right){\rm d}x_1 {\rm d}x_2\nonumber\\
		&\overset{w\to 0}{\sim}w^{-1+k_1}\int_{0}^{\Lambda}x_2^{-1+k_1+k_2}{\rm d}x_2\propto w^{-1+k_1}\;,
	\end{align}
	which indicates the divergent power is determined by numerator only,
	\begin{equation}
		k_w=k_1\;.
	\end{equation}
	
	\item Distribution for difference of two independent variates, $u=x_1-x_2$ , is
	\begin{align}
		P(u)&=\mathcal{N} \int_0^{\Lambda}x_1^{-1+k_1}x_2^{-1+k_2}\delta\left(u-x_1+x_2\right){\rm d} x_1 {\rm d} x_2\nonumber\\
		&\propto
		\begin{cases}\displaystyle
			\int_{0}^{\Lambda-u}\left(u+x_2\right)^{-1+k_1}x_2^{-1+k_2}{\rm d}x_2 & u>0\\\displaystyle
			\int_{-u}^{\Lambda}\left(u+x_2\right)^{-1+k_1}x_2^{-1+k_2}{\rm d}x_2 & u<0
		\end{cases}\nonumber\\
		&\overset{u\to 0^{\pm}}{\propto}
		\begin{cases}
			u^{-1+k_1+k_2}+\mathcal{O}\left(u^0\right) & u>0\\
			(-u)^{-1+k_1+k_2}+\mathcal{O}\left((-u)^{0}\right) & u<0
		\end{cases}\;.
	\end{align}
	Usually we only consider the case $u>0$ and this gives 
	\begin{equation}
		k_u=k_1+k_2\;,
	\end{equation}
	which is the same as summation of two independent variates.
	
\item Consider the special case, $\rho= x_1/x_2$ where $x_{1,2}$ are randomly sampled from Weibull distribution $\tilde{P}(x_i;k,l)$ with same divergent power $k$ and $l$, the result is not related to overall scale.
	\begin{align}
		P(\rho)&= \int_0^\infty \tilde{P}(x_1)\tilde{P}(x_2)\delta\left(\rho-\frac{x_1}{x_2}\right){\rm{d}}x_1{\rm{d}}x_2\nonumber\\
		&=\frac{k \rho^{-1+k}}{(1+\rho^k)^2}\overset{\rho\to 0}{\propto}\rho^{-1+k}\;,
	\end{align}
which gives us  
\begin{equation}
k_\rho=k.
\end{equation}	
\end{enumerate}

	The change of integration limit is crucial for the distribution. Here we are only interested in divergent behavior at the origin, which makes the integration much simpler to analyze. If one intends to get distribution backward (such as calculating $P(y)$ with the knowledge of $P(x)$ and $P(z=x+y)$), the above formula does not apply. Because $z$ and $x$ in this case are not independent variates.

	\section{Monte-Carlo Simulation}
	\label{appendix:mc_sim}
	
\subsection{Matrix Decomposition}	

This appendix explores the relation between distributions of mixing angles and eigenvalues decomposed from a mother matrix. Here we generate some random matrices $Y$s and decompose them to extract data of mixing angles and eigenvalues.

If we assume that matrix elements carry no weight, all elements are random variates generated independently from an intrinsic probability density function with some divergent power, $P(y)\propto y^{-1+k}$. Here we use a single flat-distributed flux parameter $a_{ij}$ to mimic the peaking behavior of distribution of matrix elements $y_{ij}$. Let
	\begin{equation}
		y_{ij}=a_{ij}^{s}\;,
		\label{eq:flux_a}
	\end{equation}
	where $a_{ij}$ is a flux parameter sampled from flat distribution $P(a_{ij})=1$ for $0\leq a_{ij}\leq 1$ and $s$ is some fixed power which is related to divergent power of $y_{ij}$ as~\cite{Sumitomo:2012wa}
	\begin{equation}
		k=\frac{1}{s}\;.
		\label{eq:yk}
	\end{equation}
The Singular Value Decomposition for mother matrix $(Y)_{ij}=y_{ij}$ gives
\begin{equation}
		Y=U_{\rm L}^{} D U_{\rm R}^{\top}\;,
\end{equation}
where $U_{\rm L,R}$ are rotation matrices expressed by its own mixing angle and $D$ is the diagonal matrix with positive eigenvalues. Here elements in $Y$ are all randomly generated from flux parameter using Eq.~\eqref{eq:flux_a} with some chosen $k$ (or equivalently $s$).

	\begin{figure}
		\centering
		\includegraphics[scale=0.9]{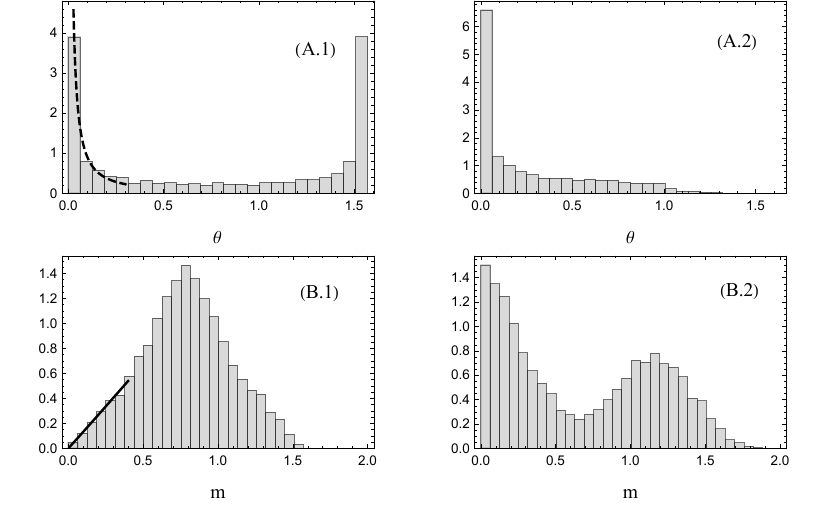}
		\caption{MC simulation of distribution of decomposition angle $\theta$ and eigenvalues $m$. Here (A) denotes the distribution generated from $k=0.2$ and (B) labels the case with $k=1.1$. Two mixing angles for each $Y$ is collected together as $\theta$ since they are fundamentally the same. Two eigenvalues $m_{1,2}$ are also put together in this figure. The solid and dashed lines in (A.1) and (B.1) are fitted functions, which mimic behavior of these distribution at the origin.}
		\label{fig:2d_mc}
	\end{figure}
	
In simplest 2D scenario, there is only one angle in left or right rotation matrix. We gather them to investigate their distributions. Figure~\ref{fig:2d_mc} contains histograms of decomposition angles and eigenvalues for real two-by-two matrix. Here the input is the randomly generated real matrix $Y$ and output is mixing angles $\theta$ and eigenvalues $m$. The plateau in part (A.2) and the valley in (B.2) is due to the fact that we are counting two eigenvalues together. Every couple of eigenvalues are not totally independent for they are decomposed from a mother matrix. There is a spike in part (B.1) of figure~\ref{fig:2d_mc}. This is expected as $y_{ij}$ is localized at value $1$ for $k>1$, which leads to a matrix with four elements approximately the same value and decomposition angle close to $\pi/4$.
	
Despite the rather complicated shape in the middle, the behaviors at the origin are quite different between A and B. The divergent powers for angle and eigenvalue are both smaller than 1 in case A and they are both larger than 1 (no divergent behavior at the origin) in case B. This gives us a hint for the reserve situation. It is reasonable to believe that when we consider angles and eigenvalues are independently sampled from their own statistical distribution. The divergent powers of them should satisfies some relations. To explore the behavior of these distributions at the origin, we pick out first few bins in these histograms and use the model Eq.~\eqref{eq:qk} to fit out divergent powers. As shown in figure~\ref{fig:2d_mc}, the dashed curve gives $P(\theta\to0)\sim 0.51 x^{-1.25}$ and the solid black curve gives $P(m\to 0)\sim 1.37 x^{1.02}$. Note that fitted dashed curve is not normalizable and one could add an exponential factor to Eq.~\eqref{eq:qk} like that in Weibull distribution to make it normalizable.
	
In 2D case, there does not exist parametrization issue. When it comes to 3D case, different parametrization would give different values of $\theta$ but with same values of $m$. Therefore, $P(\theta)$ would be different and $P(m)$ the same in different parametrization. The $k_{\rm N}$-$k_{\rm M}$ relation that we need in the main text is used for symmetric Majorana mass matrix. Here we generate three-by-three real symmetric matrices $Y_{(k)}$ and decompose them in KM and SP. Table~\ref{tab:kkkk} shows the fitted divergent powers of $k_{\rm N,SP}$, $k_{\rm N, KM}$ and $k_{\rm M}$ for each $k$ used in Eq.~\eqref{eq:yk} , where 
\begin{align*}
P(\theta_{\rm N,SP}\to 0)&\propto \theta_{\rm N,SP}^{-1+k_{\rm N, SP}}\\
P(\theta_{\rm N,KM}\to 0)&\propto \theta_{\rm N,KM}^{-1+k_{\rm N, KM}}\\
P(m\to 0)&\propto m^{-1+k_{\rm M}}\;.
\end{align*}

\begin{figure}
\centering
\includegraphics[scale=1]{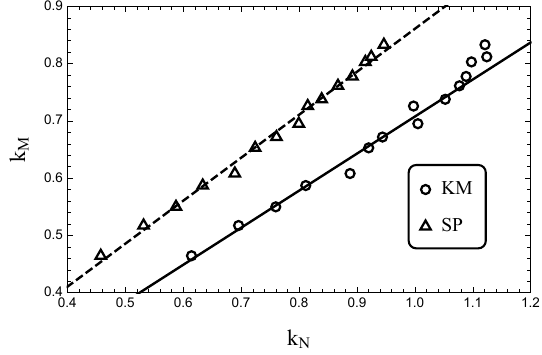}
\caption{List plot of $k_{\rm N}$ and $k_{\rm M}$ in SP and KM parametrizations. Solid line is the linear fit in KM and dashed line is that for SP.} 
\label{fig:kmkn}
\end{figure}

Note that it is difficult to generate higher $k_{\rm M}$ and $k_{\rm N}$, since higher $k$ in our simplified model indicates the effect from cut-off (see Eq.\eqref{eq:appendix_px}) at $1$ becoming more significant. The value for matrix elements are not bounded in the real world, so that high $k$ indicates a break down of this simulation, which would not generate more non-trivial $(k_{\rm M}, k_{\rm N})$ data points. Here we simply linearly extrapolate these existing data pairs and achieve the approximating numerical expression of function $h$ in Eq.~\eqref{eq:kkkk}.
\begin{align}
k_{\rm M}&\approx 0.6472\, k_{\rm N, KM} + 0.06012\nonumber\\
k_{\rm M}&\approx 0.7506\, k_{\rm N, SP} + 0.1099 \;.
\label{eq:kmkn}
\end{align}
Above different $k_{\rm N}$-$k_{\rm M}$ relations compensate the difference in different parametrization used initially.

\begin{table}
\tbl{Fitted divergent powers in different randomly generated $Y$.}
{\begin{tabular}{@{}ccccccccccc@{}}
\toprule
$k=1/s$ & $k_{\rm M}$ & $k_{\rm N, KM}$  & $k_{\rm N,SP}$\\
\colrule
$0.22$  & $0.4643$ & $0.6144$ & $0.4577$  \\
\colrule
$0.24$  & $0.5171$ & $0.6957$ & $0.5316$ \\
\colrule
$0.26$  & $0.5497$ & $0.7599$ & $0.5876$ \\
\colrule
$0.28$  & $0.5869$ & $0.8117$ & $0.6338$ \\
\colrule
$0.30$  & $0.6079$ & $0.8882$ & $0.6892$  \\
\colrule
$0.32$  & $0.6527$ & $0.9204$ & $0.7243$ \\
\colrule
$0.34$  & $0.6716$ & $0.9442$ & $0.7611$ \\
\colrule
$0.36$  & $0.6949$ & $1.005$ & $0.7997$ \\
\colrule
$0.38$  & $0.7257$ & $0.9978$ & $0.8151$ \\
\colrule
$0.40$  & $0.7375$ & $1.053$ & $0.8392$ \\
\colrule
$0.42$  & $0.7611$ & $1.077$ & $0.8673$ \\
\colrule
$0.44$  & $0.7774$ & $1.089$ & $0.8923$ \\
\colrule
$0.46$  & $0.8028$ & $1.099$ & $0.9146$ \\
\colrule
$0.48$  & $0.8117$ & $1.124$ & $0.9248$ \\
\colrule
$0.50$  & $0.8328$ & $1.121$ & $0.9466$ \\
\botrule
\end{tabular}\label{tab:kkkk}}
\end{table}

\subsection{$SO(3)$ Mixing Angles}	
 When it goes to three generations of fermions, the mixing matrix is described by a three-by-three rotation matrix, which is totally determined by three mixing angles, $\mathcal{R}_{\rm p}=\mathcal{R}_{\beta} \mathcal{R}_{\rm c}$, where
\begin{align}
\mathcal{R}_{\rm c}&=U_{\rm SP}\left(\theta_{\rm c}^{12},\theta_{\rm c}^{23},\theta_{\rm c}^{13},0\right)\nonumber\\
\mathcal{R}_{\beta}&=U_{\rm SP}\left(\beta^{12},\beta^{23},\beta^{13},0\right)\;.
\end{align} 
Here the CP-violating phases are neglected for simplicity. The algebraic relation between $\theta_{\rm p}^{ij}$ and $(\theta_{\rm c}^{ij},\beta^{ij})$ is complicated and difficult to see the relation between divergent powers for parameters. We randomly generate angles $\theta_{\rm c}^{ij}$ and $\beta^{ij}$ from distribution imitated by transformed Beta distribution defined in Eq.~\eqref{eq:beta_dist}
	\begin{align}
		P(\theta_{\rm c})&=B(\theta_{\rm c};a_{\rm c},b_{\rm c})\nonumber \\
		P(\beta)&=B(\beta;a_\beta,b_\beta)\;.
	\end{align}
	After express rotation matrix in the standard form, $\mathcal{R}_{\rm p}=U\left(\theta_{\rm p}^{12},\theta_{\rm p}^{23},\theta_{\rm p}^{13},0\right)\;$, resultant angles $\theta_{\rm p}^{ij}$ could be extracted and collected. The distribution of $\theta_{\rm p}$ resembles a transformed Beta distribution. The divergent power of $P(\theta_{\rm p})$ is approximated by the best-fit value $a$ using $B(\theta_{\rm p};a,b)$. 
	
\begin{table}
\tbl{$a_{\rm p}$ with different $a_{\rm c}$ and $a_{\beta}$ under fixed value of $b_{\rm c}=2$ and $b_{\beta}=4$.}
{\begin{tabular}{@{}ccccccccccc@{}}
\toprule
		\diagbox[width=3.7em,height=2em]{$a_{\rm c}$}{$a_{\beta}$} & $0.2$ & $0.4$ & $0.6$ & $0.8$ & $1.0$ & $1.2$ & $1.4$ & $1.6$ & $1.8$ & $2.0$\\
\colrule
		$0.2$ & $0.420$ & $0.601$ & $0.765$ & $0.899$ & $1.02$ & $1.16$ & $1.27$ & $1.37$ & $1.47$ &   $1.54$ \\
\colrule
		$0.4$ & $0.601$ & $0.741$ & $0.901$ & $1.01$ & $1.10$ & $1.21$ & $1.28$ & $1.34$ & $1.42$ &   $1.42$ \\
\colrule
		$0.6$ & $0.751$ & $0.895$ & $0.996$ & $1.09$ & $1.15$ & $1.25$ & $1.29$ & $1.33$ & $1.38$ &   $1.36$ \\
\colrule
		$0.8$ & $0.890$ & $0.991$ & $1.08$ & $1.16$ & $1.24$ & $1.28$ & $1.33$ & $1.30$ & $1.36$ &   $1.38$ \\
\colrule
		$1.0$ & $1.05$ & $1.10$ & $1.19$ & $1.25$ & $1.30$ & $1.29$ & $1.30$ & $1.31$ & $1.33$ & $1.37$   \\
\colrule
		$1.2$ & $1.16$ & $1.22$ & $1.22$ & $1.28$ & $1.30$ & $1.34$ & $1.33$ & $1.36$ & $1.36$ & $1.36$   \\
\colrule
		$1.4$ & $1.28$ & $1.29$ & $1.31$ & $1.31$ & $1.30$ & $1.34$ & $1.33$ & $1.36$ & $1.35$ & $1.35$   \\
\colrule
		$1.6$ & $1.38$ & $1.34$ & $1.34$ & $1.33$ & $1.35$ & $1.37$ & $1.33$ & $1.35$ & $1.32$ &   $1.38$ \\
\colrule
		$1.8$ & $1.50$ & $1.39$ & $1.37$ & $1.33$ & $1.33$ & $1.36$ & $1.35$ & $1.38$ & $1.40$ &   $1.38$ \\
\colrule
		$2.0$ & $1.51$ & $1.42$ & $1.36$ & $1.37$ & $1.35$ & $1.34$ & $1.34$ & $1.35$ & $1.41$ & $1.42$   \\
\botrule
\end{tabular}\label{tab:3d_mc}}
\end{table}
	
	Here the value of $b_{\rm c}$ and $b_{\beta}$ basically has no impact on $a_{\rm p}$. So we take specific values of $b_j$s to investigate the relation between $a_j$s. The following expression provides a good fit of the data in table~\ref{tab:3d_mc},
	\begin{equation}
		a_{\rm p}\approx 0.990\left(a_{\rm c}+a_{\beta}\right)-0.175\left(a_{\rm c}+a_{\beta}\right)^2 + 0.0427\left(a_{\rm c}-a_{\beta}\right)^2\;.
	\end{equation}
	$a_{\rm c}$ and $a_{\beta}$ are approximately symmetric from numerical evaluation. Therefore, there is no $(a_{\rm c}-a_{\beta})^1$-term. The constant term is also absent here due to the fact that $\theta_{\rm p}$ is generated from two random variables and there should not exist some intrinsic independent part in its divergent power. The coefficient of linear sum dominates for $a_j\sim\mathcal{O}(1)$. This means that the even in $SO(3)$ case, the divergent power relation derived in $O(2)$ toy model Eq.~\eqref{eq:kpcb} still provides a good approximation.

\bibliographystyle{ws-ijmpa}
\bibliography{ref}

\end{document}